# Programmable Magnetic Hysteresis in Orthogonally-Twisted Two-Dimensional CrSBr Magnets via Stacking Engineering

*Carla Boix-Constant, Andrey Rybakov, Clara Miranda-Pérez, Gabriel Martínez-Carracedo, Jaime Ferrer, Samuel Mañas-Valero\*, Eugenio Coronado\**

C. Boix-Constant, A. Rybakov, C. Miranda-Pérez, E. Coronado

Instituto de Ciencia Molecular (ICMol), Universitat de València, Catedrático José Beltrán 2, Paterna, 46980 Spain.

E-mail: eugenio.coronado@uv.es

G. Martínez-Carracedo, J. Ferrer

Departamento de Física, Universidad de Oviedo, 33007 Oviedo, Spain.

Centro de Investigación en Nanomateriales y Nanotecnología, Universidad de Oviedo-CSIC, 33940 El Entrego, Spain

S. Mañas-Valero

Department of Quantum Nanoscience, Kavli Institute of Nanoscience,

Delft University of Technology, Delft 2628CJ, The Netherlands.

E-mail: S.ManasValero@tudelft.nl

Twisting two-dimensional van der Waals magnets allows the formation and control of different spin-textures, as skyrmions or magnetic domains. Beyond the rotation angle, different spin reversal processes can be engineered by increasing the number of magnetic layers forming the twisted van der Waals heterostructure. Here, we consider pristine monolayers and bilayers of the A-type antiferromagnet CrSBr as building blocks. By rotating 90 degrees these units, we fabricate symmetric (monolayer/monolayer and bilayer/bilayer) and asymmetric (monolayer/bilayer) heterostructures. The magneto-transport properties reveal the appearance of magnetic hysteresis, which is highly dependent upon the magnitude and direction of the applied magnetic field and is determined not only by the twist-angle but also by the number of layers forming the stack. This high tunability allows switching between volatile and non-volatile magnetic memory at zero-field and controlling the appearance of abrupt magnetic reversal processes at either negative or positive field values on demand. The phenomenology is rationalized based on the different spin-switching processes occurring in the layers, as supported by micromagnetic simulations. Our results highlight the combination between twist-angle and number of layers as key elements for engineering spin-switching reversals in twisted magnets, of interest towards the miniaturization of spintronic devices and realizing novel spin textures.

**1. Introduction**

Van der Waals (vdW) magnets are attracting increasing interest among the two-dimensional (2D) materials due to their emergent properties in the 2D limit and potential applications in areas such spintronics, including magnetic memories and sensing, or magnonics.[1] In particular, A-type antiferromagnets (like $CrI_3$, CrSBr or $CrPS_4$, among others) are formed by ferromagnetic layers weakly coupled antiferromagnetically.[2–4] The stacking of two of these layers mimics the minimal structure of a conventional spin-valve, being the vdW gap the spacer between the ferromagnetic layers, where high/low resistance values are reached depending on the layers´ antiparallel/parallel magnetization configuration, which can be controlled by applying an external magnetic field.[5–9] Since vdW magnets can be exfoliated down to the single layer limit, as commonly performed in the field of 2D materials, the bilayer scenario represents the thinnest spin-valve-like case. [5–7]

An attractive possibility offered by 2D vdW materials is that of creating twisted heterostructures. Beyond the pioneering experiments on magic-angle twisted bilayer graphene –where superconductivity emerges by a twist angle of 1.1º–, the twistronics field has rapidly expanded to investigate other 2D materials, providing a unique platform to develop moiré physics with optical, electrical or magnetic properties.[10,11] Still, the study of twisted 2D magnetic heterostructures is very recent and has focused mainly in $CrI_3$, a van der Waals magnet with an out-of-plane uniaxial spin anisotropy.[12–15] Indeed, spin-anisotropy is a key ingredient in twisted 2D magnets which, in some cases, can be even more relevant than the formation of moiré superlattices, as we have shown for the case of orthogonally-twisted CrSBr ferromagnetic monolayers.[16] CrSBr is a magnetic semiconductor that has gained recent interest due to the coupling between its electrical, optical and magnetic properties,[17–22] ordering at a relative high temperature (*ca.* 140 K)[6,23] and with highly anisotropic magnetic properties, exhibiting an in-plane uniaxial spin anisotropy along the *b* axis (easy-magnetic axis).[24] The twisted-CrSBr vdW spin valves have provided the opportunity to study the emergent properties arising from the competition between in-plane uniaxial spin anisotropy, interlayer antiferromagnetic interactions and the direction of the magnetic applied field. In this regard, we observed a multiple spin switching with hysteresis in the magneto-resistance (MR) measurements as a result of the interplay between intra-layer spin reorientation and interlayer spin-flip processes when the field is applied along the easy axis of one of the layers, highlighting the appearance of non-volatile magnetic memory at zero-field triggered by the application of the proper magnetization protocol.[16] In a subsequent recent work, Parkin´s group has investigated the influence of the angle in twisted CrSBr layers by varying the angle from 0º to 90º.[25] Interestingly, non-volatile giant tunnelling MR is observed at zero field for intermediate twist angles (35º) in the all-antiferromagnetic bilayer/bilayer case, being of potential interest for magnetic memories.[25] Recently, Healey et al. have imaged orthogonally-twisted CrSBr heterostructures (monolayer/monolayer, bilayer/bilayer and bilayer/trilayer) by nitrogen-vacancy centers in diamond microscopy, observing the formation and propagation of magnetic domains.[26] In parallel, theoretical calculations on twisted magnetic heterostructures have predicted the possibility of reaching multiferroics, controllable magnetic domains and even more complex topological spin textures, as merons or skyrmion bubbles.[27–30]

Beyond the twist-angle, the number of stacked pristine and twisted magnetic layers is as well highly-relevant for engineering the spin-switching reversal processes. In this work we exploit this concept by increasing the number of pristine CrSBr layers, while fixing a twist-angle of 90 degrees. In particular, we employ pristine monolayers and bilayers as building blocks for fabricating orthogonally-twisted monolayer/monolayer, monolayer/bilayer and bilayer/bilayer CrSBr devices. By tracking their MR properties, we show the high tunability of the magnetization reversal in orthogonally-twisted heterostructures, that can be controlled by selecting the number of layers of the vdW heterostructure and the direction and magnitude of the applied magnetic field, hence allowing the switching between volatile and non-volatile memory at zero-field. Based on micromagnetic simulations, our experimental observations can be rationalized as a competition between the spin-switching mechanism occurring in the different layers (spin-flip and spin-reorientation) and the applied external magnetic field. Overall, this high tunability paves future directions for engineering the properties of atomically-thin spin-valve devices based on vdW magnets by selecting not only the twist-angle but, as well, the number of pristine and twisted layers stacked in the magnetic vdW heterostructure.

## 2. Results and discussion

Monolayers and bilayers of CrSBr are mechanically exfoliated from their bulk counterpart under inert conditions and integrated into vertical van der Waals heterostructures for inspecting the magneto-transport properties. A representative orthogonally-twisted CrSBr vdWs heterostructure is shown in **Figure 1.a-b**, formed by a monolayer deposited on top of a bilayer of CrSBr with their respective easy-magnetic axis forming an in-plane angle of 90°. This heterostructure is located between few-layers graphene which are on top of pre-lithographed electrodes. The whole device is encapsulated with h-BN (see **Methods** and **Supporting Information Section 1** for further details). Note that in this orthogonal configuration, the application of a magnetic field along the easy-axis of one building unit implies that the field is oriented along the intermediate-axis of the other one.

First, we summarize the magneto-transport properties of the pristine monolayer and bilayer case (**Figure 1c,d**). As previously reported,[5,6] the pristine monolayer device exhibits no resistance dependence for fields pointing along the easy magnetic axis (*b*) and a resistance enhancement for fields pointing along the intermediate magnetic axis (*a*), that saturates at ± 1 T, as shown in **Figure 1.c**. On the contrary, in agreement with a spin-valve picture, the pristine bilayer (**Figure 1.d**) exhibits a sharp resistance drop at ± 0.2 T for fields along the *b* axis and a smoother reduction for fields along the *a* axis, saturating at ± 1 T. This magnetic behaviour corresponds to the different spin-switching mechanisms occurring in CrSBr: a pure spin-flip of the layers for fields applied along the easy-axis (*b*) and a combination of spin-reorientation and spin-flip if the fields are applied along the intermediate (*a*) or hard (*c*) magnetic axis.[5,6]

Next, we turn our attention to the orthogonally-twisted cases, discussing the monolayer/monolayer (**Figure 1.e**), monolayer/bilayer (**Figure 1.f**) and bilayer/bilayer (**Figure 1.g**) case at 2 K, where the field is applied along the easy/intermediate magnetic axis of the bottom/top layer (θ = 90° in the sketch of **Figure 1.b**). Overall, the three orthogonally-twisted cases exhibit a similar behaviour at large applied fields (**Figure S1**), reassembling the pristine bilayer one (**Figure 1.d**): a positive MR that decreases down to zero for fields with modulus larger than 1T. The MR is defined as $100 \cdot [R(B) – R(P)]/R(P)$, being R(P) the resistance when the magnetization of all the layers is parallel (in our case, we consider the resistance value at 3 T). In contrast, the field-dependences are extremely dissimilar at smaller fields (**Figure 1.e-g**), where diverse hysteretic behaviours are observed. For example, while sweeping from negative to positive fields (red trace in **Figure 1.e-g**), the monolayer/monolayer exhibits a gradual trend, that is in stark contrast with the monolayer/bilayer and bilayer/bilayer cases, where sharp resistance jumps are observed. This indicates that, in the monolayer/monolayer heterostructure, the MR behaviour is dominated by a smooth spin-reorientation process while, when a pristine bilayer is involved, the abrupt spin-flip process dominates. All the heterostructure show hysteretic behaviour. This hysteresis can be better visualized by computing the magnitude

ΔX = X$_{+B→-B}$ − X$_{-B→+B}$, where X stands for either R or MR. Thus, non-zero ΔX values indicate hysteretic response, as indicated by the yellow traces in **Figure 1.e-g**. Note that the appearance of an hysteresis response contrasts with the pristine cases, where no significant hysteresis is observed, as shown in **Figure 1.c-d**.[5,6] The overall phenomenology is robust and consistent across the 11 orthogonally-twisted devices studied in this work (see **Methods** and **Supporting Information Section 1**).

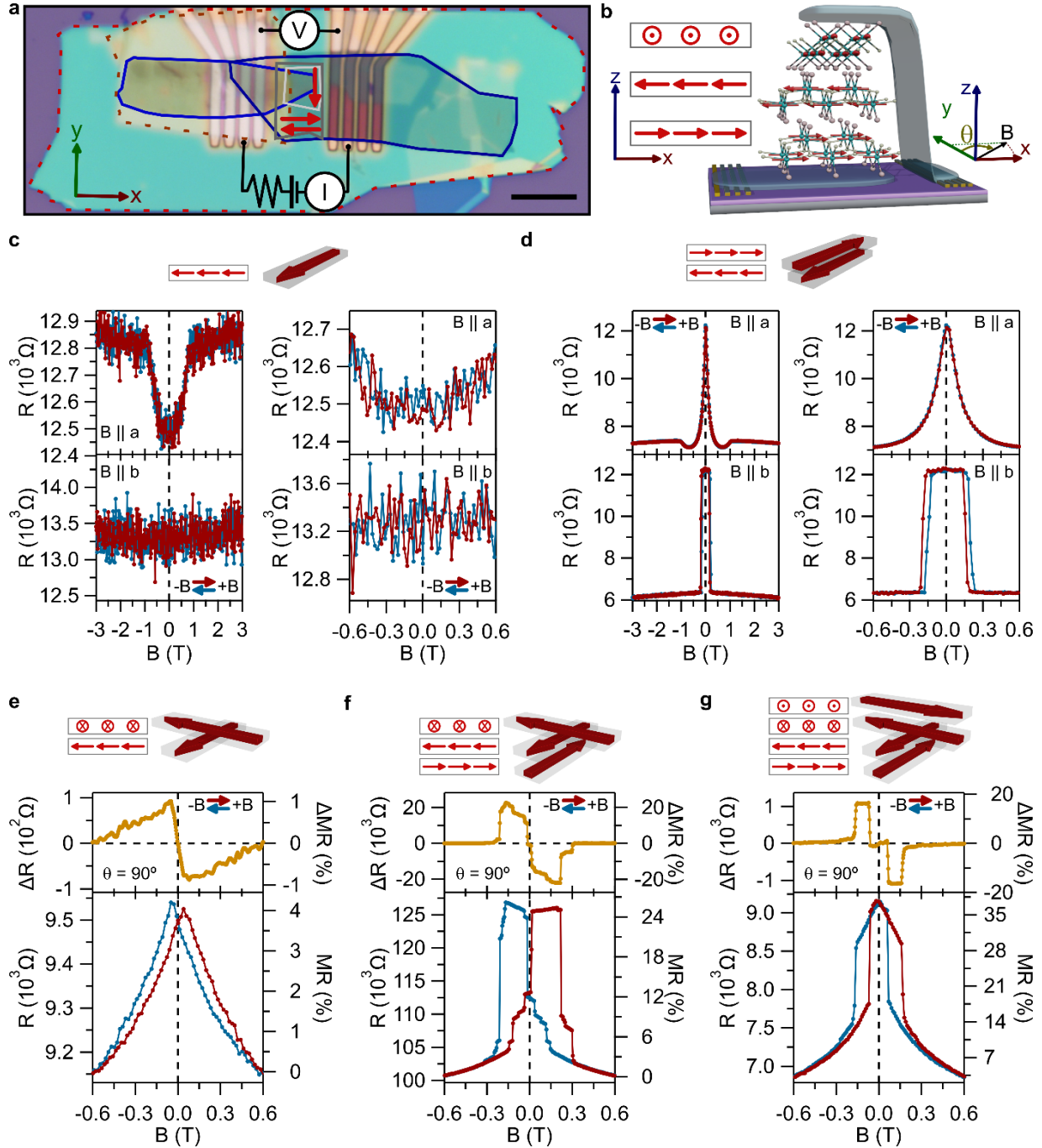

**Figure 1.- Magneto-resistance properties of orthogonally-twisted monolayer/monolayer, monolayer/bilayer and bilayer/bilayer CrSBr heterostructures.** a) Optical image of a device comprising a monolayer (grey line with one red arrow enclosed) and bilayer (dark grey line with two red arrows enclosed) CrSBr, whose *ab* axes are

orthogonal, placed in between few-layers graphene (blue lines). Different insulating hexagonal boron nitride layers (red dashed lines) are used both to avoid shortcuts and to protect the heterostructure. The red arrows indicate the easy magnetic axis (*b*) of each CrSBr building block, being the intermediate magnetic axis (*a*) perpendicular to it. The hard magnetic axis (*c*) corresponds to the out-of-plane direction. The electrical measurement scheme is sketched, where I refers to current and V to voltage. Scale bar: 5 µm. b) Representation of an orthogonally-twisted monolayer/bilayer CrSBr heterostructure (not to scale) deposited on top of pre-patterned gold electrodes. Few-layers graphene (blue layers) are enclosing the twisted CrSBr layers (pink, yellow and cyan balls correspond to bromine, sulphur and chromium atoms, respectively). Inset: Schematic view of the bottom bilayer and top monolayer. The red arrows represent the spin lying along the easy magnetic axis, assuming negligible interlayer magnetic interactions. c-d) Field dependence of the resistance for the pristine monolayer (c) and pristine bilayer (d) for fields applied along the *a* (top panel) and *b* (bottom panel) axis. e-g) R and MR field dependence and its increment ($\Delta X$), defined as $\Delta X = X_{+B \to -B} - X_{-B \to +B}$ (X indicates either the R or the MR) for e) monolayer/monolayer, f) monolayer/bilayer and g) bilayer/bilayer orthogonally-twisted CrSBr heterostructures at T = 2 K for a field-angle of $\theta = 90°$, as sketched in b. Sweeping up (down) trace are denoted in red (blue).

An interesting observation refers to the tunability of the spin switching achieved in these heterostructures. In the pristine bilayer, when B is oriented along the *b* axis, the field-dependent MR curves are symmetrical and show a high-resistance plateau around zero-field followed by a sudden drop at ±0.2 T (see **Figure 1.d**). These field values correspond to the Zeeman energy required for performing the spin-flip. In contrast, in the orthogonally-twisted monolayer/bilayer (**Figure 1.f**) and bilayer/bilayer (**Figure 1.g**) heterostructures, the resistance drops are placed asymmetrically with respect to the sign of B. Indeed, the monolayer/bilayer heterostructure shows an extreme behaviour where the full resistance plateau and drops appear shifted towards positive magnetic fields when sweeping up B (red trace in **Figure 1.f**). In particular, these drops occur at +0.01 T and +0.22 T. The behaviour is reversed when B is swept down (blue trace in **Figure 1.f**). The asymmetry is less pronounced in the bilayer/bilayer heterostructure since the resistance drops occur at a negative field of -0.06 T and at a positive field of +0.17 T for the field-increasing trace (see red curve in **Figure 1.g**), with a fully reversed behaviour for the retrace path (see blue curve in **Figure 1.g**).

The zero-field resistance value is the same while sweeping up and down the magnetic field for the three types of orthogonally-twisted heterostructures studied here, as can be seen in **Figure 1.e-g**, thus displaying volatile memory behaviour. To further investigate the irreversibility processes underlying the spin switching mechanism that occur on these heterostructures, we perform first order reversal curves (**Figure 2**).[31,32] We saturate the magnetization at negative fields (-3 T), decrease B to -0.4 T and, then, execute the magnetic sequence $-0.4\ T \to B_{max} \to -0.4\ T$, where $B_{max}$ is increased in 20 mT steps from -0.4T to +0.4T (see **Supporting Video 1-3**). Selected $B_{max}$ values are shown in **Figure 2** while the full dataset is presented in the **Supporting Information Section 2**. For the monolayer/monolayer heterostructure, the resistance exhibits a positive slope with no hysteresis for $B_{max} < 0.08$ T, as indicated in **Figure 2.a**. Above this threshold, the trace (red curve in **Figure 2.a**) and retrace (blue curve in **Figure 2.b**) differ around zero field, thus marking the appearance of hysteresis (non-zero ΔR values at zero-field) and, therefore, a non-volatile memory effect. For $B_{max} > 0.58$ T, the curves are symmetric as shown in **Figure 1.e**. A striking difference is observed in the monolayer/bilayer (**Figure 2.b**) and bilayer/bilayer (**Figure 2.c**) case if compared with the monolayer/monolayer (**Figure 2.a**) scenario since the hysteretic effects start appearing at negative values of $B_{max}$. Of special interest is the phenomenology observed in the monolayer/bilayer heterostructure for $B_{max}$ values between 20 mT and 220 mT, where a hysteresis effect appears even at zero-field. The bilayer/bilayer show no hysteresis at zero field for any $B_{max}$ value. Applying the reverse sequence (+0.4 T → $B_{max}$ → +0.4 T, where $B_{max}$ is increased in 20 mT steps from +0.4 T to -0.4 T) delivers a mirror image with respect the R/MR axis (see **Supporting Video 1-3** and **Supporting Information Section 2**). Overall, the combination of applied magnetic field and number of layers forming the orthogonally-twisted heterostructure allows for a selective switch between non-hysteretic and hysteretic states at zero-field, thus triggering between volatile and non-volatile memory behaviours.

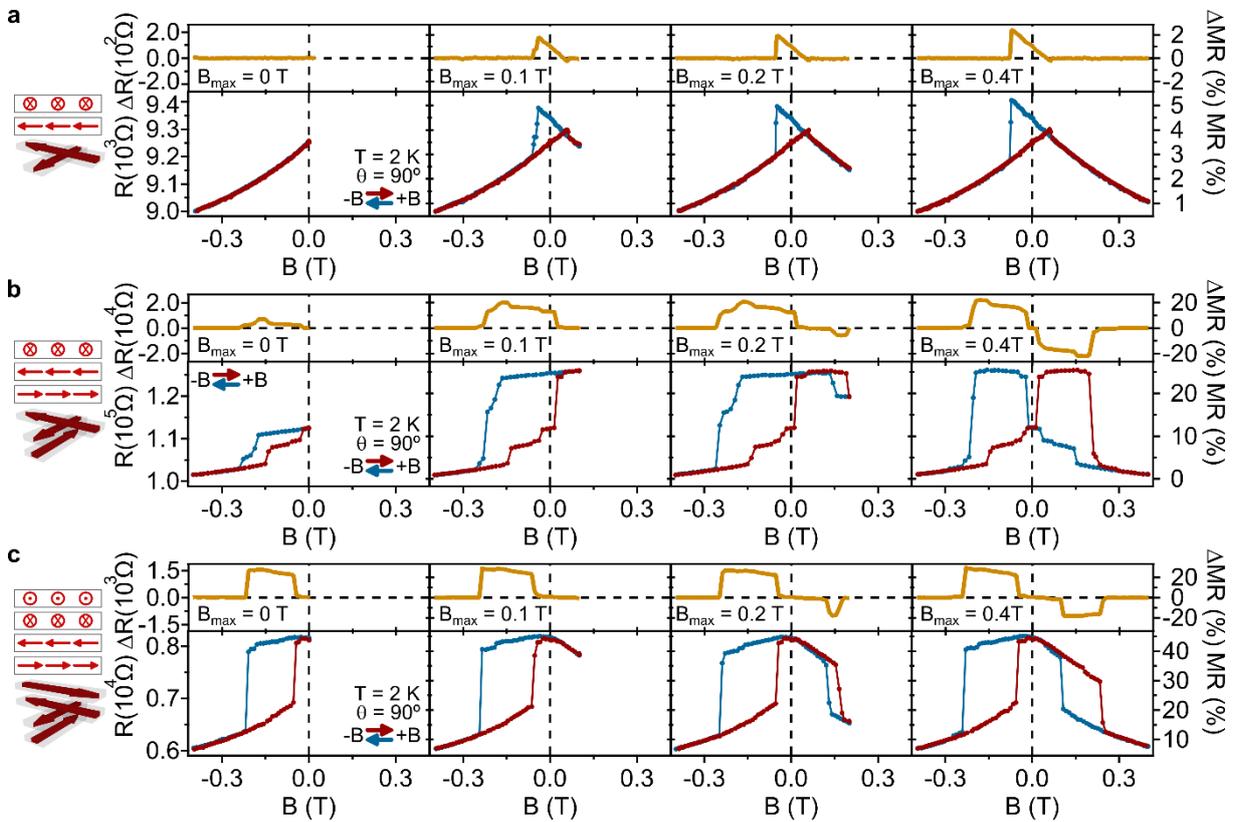

**Figure 2.- Volatility switching in orthogonally-twisted CrSBr.** First order reversal curves (FORCs) for orthogonally-twisted **a**) monolayer/monolayer, **b**) monolayer/bilayer and **c**) bilayer/bilayer CrSBr heterostructures considering the field sequence $-0.4\,\text{T} \rightarrow B_{max} \rightarrow -0.4\,\text{T}$. $B_{max}$ is incremented sequentially in steps of 20 mT. Selected curves are shown in the panels. Sweeping up (down) trace is denoted in red (blue).

The magnetization switching mechanism of the orthogonally-twisted layers is governed by the competition between two spin switching mechanisms: spin-flip versus spin reorientation. This competition renders a strong dependence on the spin-anisotropy and, therefore, on the relative orientation between the magnetic field direction and the easy-magnetic axis of each of the two orthogonally-twisted buildings blocks involved. As a result, the hysteresis curves exhibit a wide variety of switching fields depending on the direction of the applied field, as shown in **Figure 3**.**a-c** for selected angles (the complete angular dependence is shown in the **Supporting Information Section 2**). The monolayer/monolayer and bilayer/bilayer heterostructures, shown in **Figure 3.a** and **Figure 3.c**, respectively, represent symmetric situations with respect to the orientation of the magnetic field. Thus, both $\theta = 0°$ and $\theta = 90°$ orientations should be equivalent since the magnetic field is applied along the easy magnetic axis of one building block and along the intermediate-magnetic axis of the other one. However, even though the general trends are similar at $\theta = 0°$ and $\theta = 90°$, we note that the experimental curves are not identical. The differences can be attributed to either slight field misalignments, to fabrication imperfections in the twist angle or to different strains in the layers. Indeed, strain can modulate the interlayer magnetic exchange[33] or act as a source of magnetic domain nucleation, as recently pointed by squid-on-a-tip and diamond magnetometry.[26,34] A more intriguing configuration is offered by the monolayer/bilayer heterostructure (**Figure 3.b**), where a clear asymmetric configuration with respect the orientation of the external field at $\theta = 0°$ and $\theta = 90°$ is observed. The full angular dependence of the three heterostructures is summarized in the ΔMR plots in **Figure 3.d-f**, that highlight how the magnetization switching is extremely-dependent on the field direction. Indeed, the switching field values and hysteresis (ΔMR) can be tuned in the range ±0.5 T by simply selecting the number of layers in each building block together with the orientation and strength of the applied field (**Figure 3.e-f**). We note that this angular dependence is robust, as evidenced by the reproducibility of the trends throughout multiple devices (**Supporting Information Section 1.1**). The high sensitivity of the MR on the field direction could be potentially employed for sensing not only the magnitude but also the direction of the magnetic field.

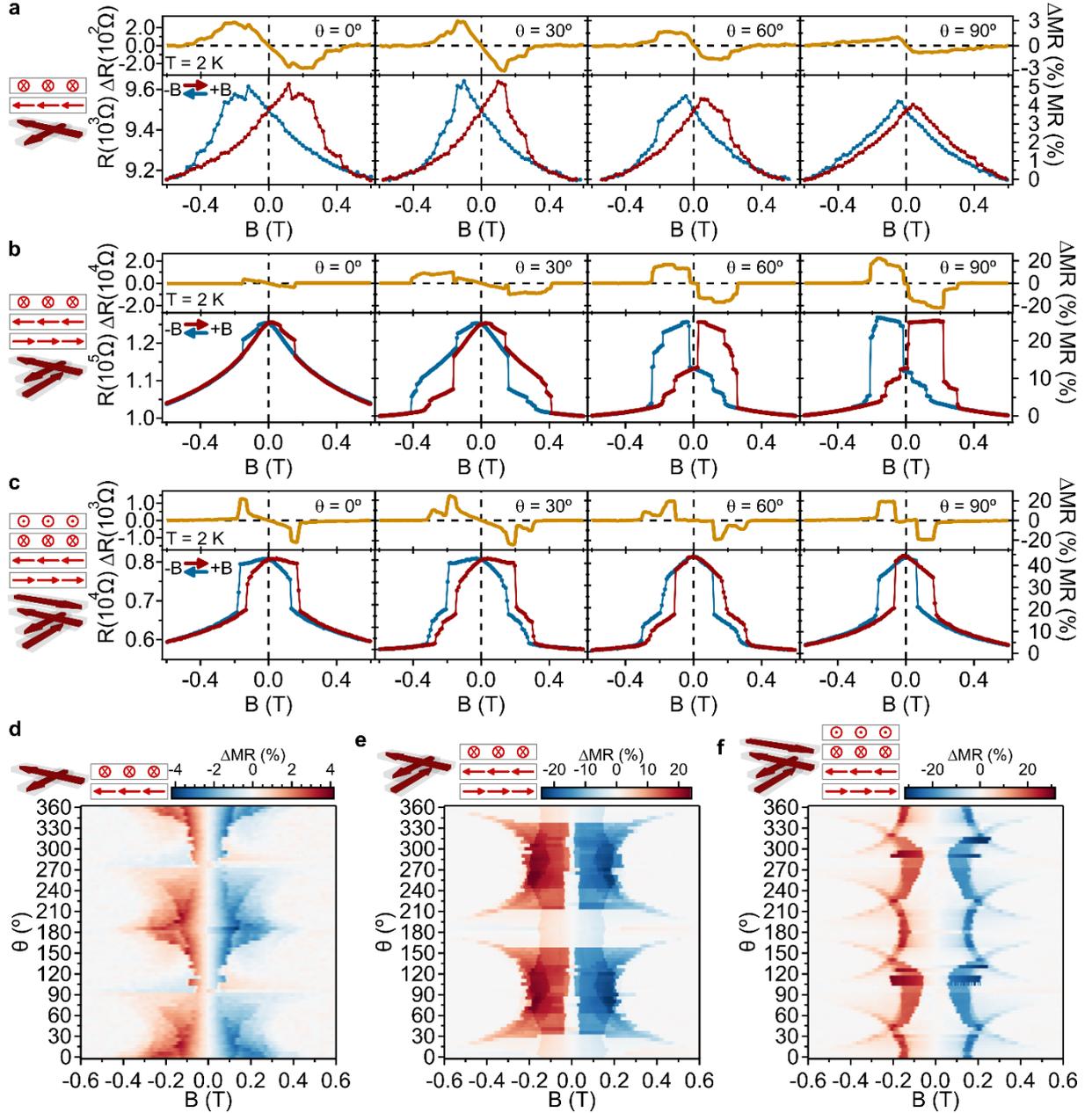

**Figure 3.- Magnetic switching in orthogonally-twisted CrSBr triggered by the field direction.** a-c) Magnetic-switching at selected angles and d-e) two-dimensional ΔMR plots for orthogonally-twisted a,d) monolayer/monolayer, b,e) monolayer/bilayer and c,f) bilayer/bilayer CrSBr. Sweeping up (down) trace is denoted in red (blue). MR is defined as MR (%) = 100·[R(B) – R(P)]/R(P), where R(P) is the resistance in the parallel state (in this case, at 3 T). Increments (Δ) are defined as $\Delta X = X_{+B \rightarrow -B} - X_{-B \rightarrow +B}$, where X is either resistance (R) or magneto-resistance (MR).

Finally, we analyse both the field and temperature dependence of the three types of heterostructures while sweeping the field from negative/positive to positive/negative values (left/middle panel in **Figure 4**). We observe that the temperature dependence is similar to the pristine CrSBr case for the three types,[5] apart from the emergence of hysteresis, shown in the right panel in **Figure 4**. The hysteretic behaviour is highly dependent on the number of layers and on the orientation of the applied field direction, as discussed above. The MR response is negligible above 200 K, then develops below 200 K as a consequence of the onset of short-range correlations, however showing no hysteresis.[5,35,36] The hysteretic effects appears at $T_c \sim$ 150 K, and increase upon further cooling down. Temperature and field dependence at different applied field directions are shown in the **Supporting Information Section 1.3.**, being the main differences between them related to the hysteresis switching fields, as discussed above, but not in the thermal behaviour.

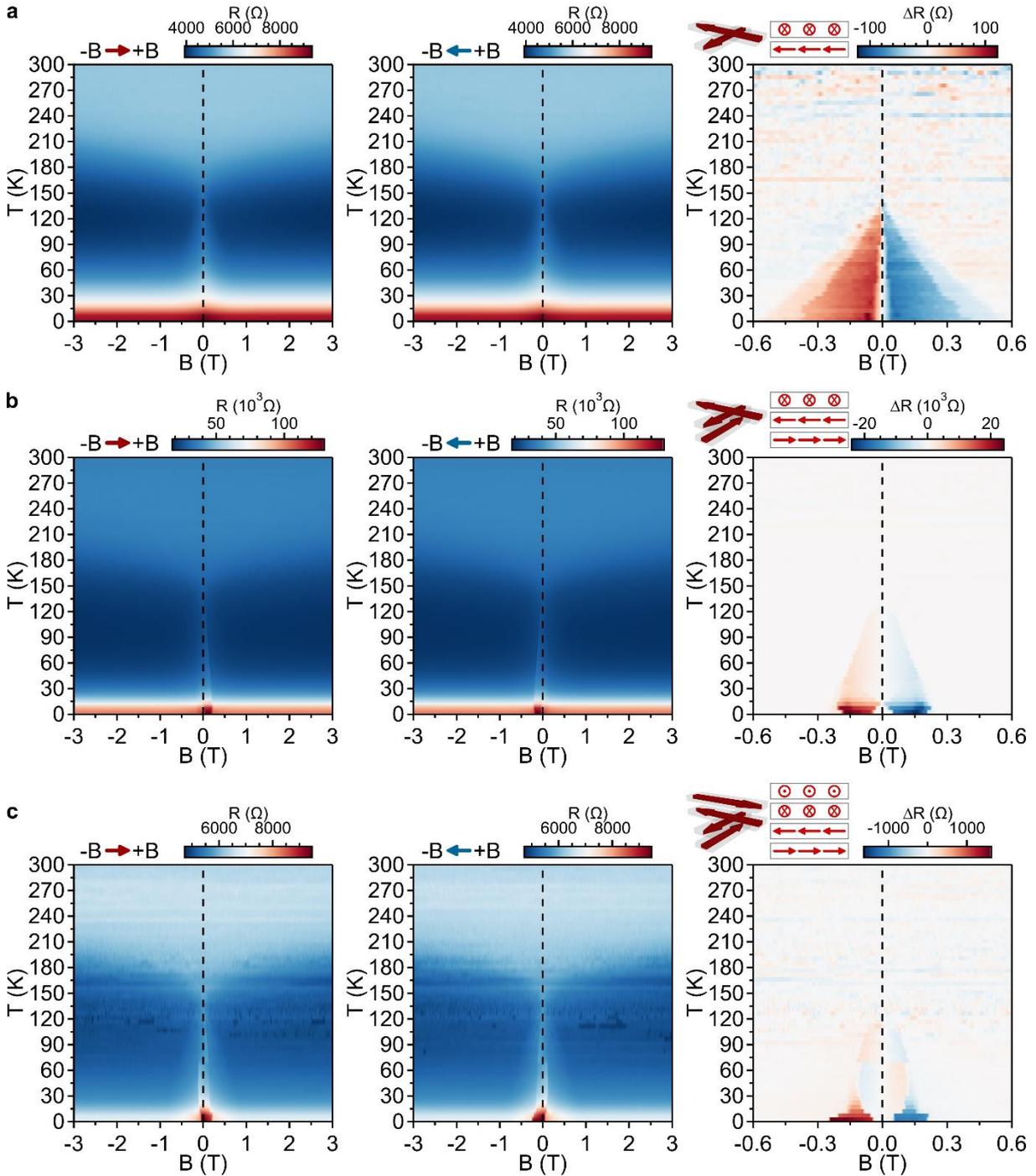

**Figure 4.- Temperature dependence of the magnetism in orthogonally-twisted CrSBr.** Resistance dependence while sweeping the field from negative to positive (left panel), from positive to negative (middle panel) as well as its difference ΔR (right panel) for **a**) monolayer/monolayer, **b**) monolayer/bilayer and **c**) bilayer/bilayer orthogonally-twisted CrSBr heterostructures.

For unravelling the relevant spin-switching mechanism taking place on the three types of orthogonally-twisted heterostructures, we have performed micromagnetic simulations.[37,38] First, we have simulated the pristine CrSBr bilayer subject to an external magnetic field oriented along its easy axis. We have found that a careful parametrisation of the effective easy-axis anisotropy $K = K_b - K_a$ with respect to the interlayer exchange coupling (J) is required to obtain robust results. For small anisotropy, defined by the ratio K/J, the main spin-switching mechanisms are spin-flop processes. However, the spin-flop behaviour is suppressed above a K/J threshold so that only an intermediate antiferromagnetic phase appears, where the relevant spin-switching mechanisms are spin-flip processes (see **Figures S19.a-b**). We find that the magnetic behaviour of CrSBr corresponds to this second scenario.[5]

We have simulated next the orthogonally-twisted heterostructures. In analogy with the experimental devices, the simulation consists of two partially overlapping square blocks that lay one on top of the other, that are rotated by 90°, as illustrated in **Figure 5.a**. Each block contains either a monolayer or a bilayer with the easy magnetic axis of the top and bottom blocks lying along the *y* and *x* directions, respectively. An example of the magnetic axes disposition relative to the *xyz* coordinates is shown in **Figure 5.b** for the monolayer/bilayer heterostructure, where the monolayer block is located on top of the bilayer block. We define for each of the possible layers an in-plane angle $\varphi_i$ between the direction of its magnetization and the *x* axis. We note that the distribution of angles $\varphi_i$ across the device can be linked with the MR signal.[39] Considering the orthogonally-twisted heterostructures, we denote by α or β the angle subtended by the total magnetization whenever two layers are untwisted or twisted, respectively. Then in the monolayer/monolayer heterostructure, we define α for the untwisted case and β for the orthogonally-twisted heterostructure. For the monolayer/bilayer heterostructure, shown in **Figure 5.b,** we define an angle α between the two layers in the bilayer block, and an angle β between the upper layer in the bilayer block and the monolayer layer. For the bilayer/bilayer heterostructures, we define the angles α and α´ for the two layers at the bottom and top bilayer blocks, and the angle β for the two middle layers, that each belong to a different block, as shown in **Figure S23**. Thus, the spin-flip process in the untwisted bilayer corresponds to a switch from α = 180° (antiparallel magnetization of the layers) to α = 0° (field-induced ferromagnetic-like state with parallel magnetization of the layers). We show in the **Supporting Videos 4-7** simulations of dynamics of the pristine bilayer as well as the orthogonally-twisted monolayer/monolayer, monolayer/bilayer and bilayer/bilayer heterostructures while sweeping the external magnetic field modulus and direction ($\theta = 0°$, 30°, 60° and 90°; further angles are complementary due to symmetry). Note that the monolayer/monolayer and bilayer/bilayer heterostructures possess $C_4$ symmetry around the *z* axis, thus a $\pi/2$ periodicity under the rotation of an in-plane magnetic field. In contrast, the monolayer/bilayer heterostructures exhibit $C_2$ symmetry, thus a $\pi$ periodicity, that agrees well with our experiments (see **Figure 3.d-f** and **Supporting Information Section 1.1**). As an illustration, we show in **Figure 5.c** different snapshots of the dynamics of the monolayer/bilayer heterostructure while sweeping the magnetic field along the *x* direction. The arrows indicate the local direction of the magnetization for the different layers while the colours indicate the angle $\varphi_i$ measured from the *x* axis for each layer. Based on the simulations for the different heterostructures, two main spin switching mechanisms can be clearly identified: a gradual spin-reorientation for fields pointing along the intermediate magnetic axis (see, as an example, the top layer in **Figure 5.c**) and a sudden spin-flip reversal for fields pointing along the easy-magnetic axis (middle and bottom layer in **Figure 5.c**). The interplay between the two mechanism leads to the formation of a canted region in the overlap area, which forms a magnetic domain, denoted with a different colour tone in **Figure 5.c**.

We plot in **Figures 5.e-f** the α and β angles obtained for each heterostructure in our simulations as a function of the external magnetic field. These plots follow well the overall trends observed experimentally. Two main features can be highlighted in the simulated spin switching process: sharp steps originated by spin-flip processes, as well as smoother evolutions due to gradual spin-reorientation processes. Overall, our simulations support the notion that control of the switching fields of the heterostructure can be achieved by changing the number of layers in each block even for a fixed twist-angle. We find that varying this number produces an effect similar to that of increasing the K/J ratio in the simpler bilayer heterostructure. Therefore, the addition of an extra orthogonally-twisted layer is proposed here to be an alternative to tuning the interlayer exchange in the bilayer itself.

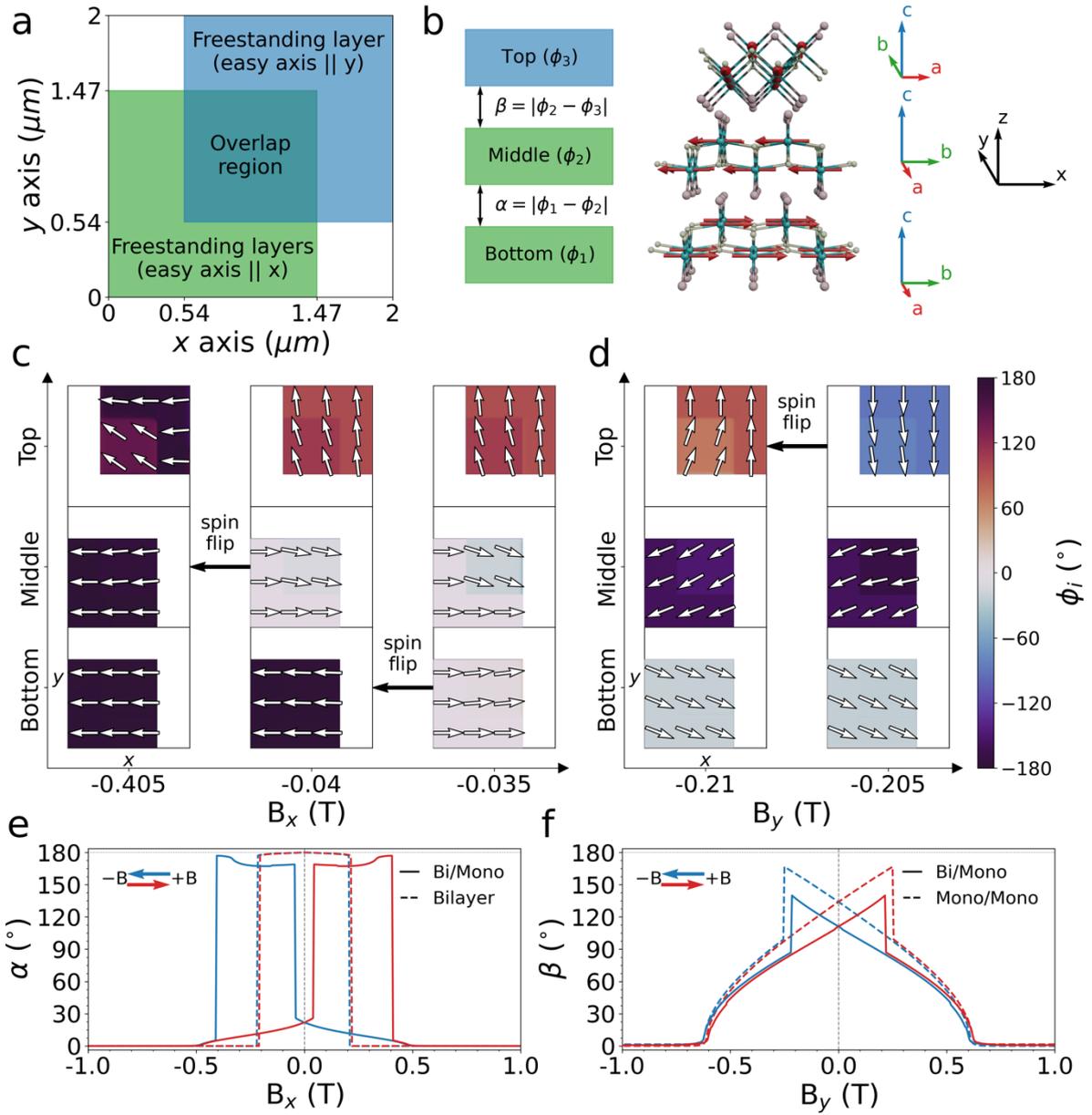

**Figure 5. Micromagnetic simulations of orthogonally-twisted CrSBr. a)** Schematic top view of the simulated twisted devices where two square blocks of layers are placed one on top of the other, rotated by 90º. The top block has its easy axis (*b*) lying in the y-direction and is drawn in blue colour; the bottom block has its easy axis lying in the x-direction; the overlap area between the top block is indicated by the darker square at the center. **b)** Example of the magnetic axes disposition with respect the *xyz* coordinates used in the simulations for the monolayer/bilayer case **c-d)** Snapshots of micromagnetic simulations for the monolayer/bilayer device (**Supporting Video 6**) with the magnetic field applied along the direction of the **c)** bilayer's easy axis (x) and **d)** monolayer's easy axis (y). **e)** Angle α between the magnetization direction of the two layers of the pristine bilayer part under the sweep of magnetic field oriented along the easy axis of bilayer. **f)** Angle β between the two monolayers rotated by 90° under the sweep of the magnetic field oriented along the easy axis of one of the monolayers (for the monolayer/monolayer device) and along the easy axis of the monolayer for the monolayer/bilayer device. In **e-f)**, the magnetic field sweep from left to the right is a mirror image of the magnetic field sweep from right to the left and is plotted for illustrative purposes.

In short, the phenomenology exhibited by the orthogonally-twisted heterostructures is rationalized based on the different spin-switching mechanisms of CrSBr, that is, spin-flip and spin-reorientation of the layer's magnetization for fields along the easy and intermediate magnetic axis, respectively. Whereas the spin-reorientation is a continuous process, the spin-flip triggers an abrupt change of the magnetic configuration. Therefore, while sweeping from negative to positive values (red trace in **Figure 3**), the twisted monolayer/monolayer heterostructure is characterized by a resistance maximum at positive field values originated by a spin-flip process, and by no-switching at negative field values. In contrast, in the bilayer/bilayer heterostructure, a first enhancement of the resistance is observed at negative values, followed by a sharp decrease at positive values, as expected for the spin-flip of the pristine bilayer. Regarding the monolayer/bilayer case (**Figure 3.b**), at $\theta = 0°$ the field is applied along the easy magnetic axis of the monolayer and, consequently, the spin-flip of the monolayer yields to a switch only at positive values while sweeping from negative to positive fields (red trace in **Figure 3.b**). On the other hand, at $\theta = 90°$, the field is aligned along the easy-magnetic axis of the pristine bilayer, being the spin-flip of the bilayer the most relevant spin-switching mechanism, which turns out to sharp resistance changes both at negative and positive fields. At intermediate angles, where the field is not aligned to any easy-magnetic axis, the interpretation of the spin-switching mechanism is more complex, since it depends on a delicate interplay between the spin-flip and spin-reorientation of the layers with respect to the applied magnetic field (Zeeman energy), being the formation of spatial magnetic inhomogeneities —as magnetic domains— more likely, in line with the experimental observation of multistep transitions (**Figure 2**) and micromagnetic simulations (**Figure 5** and **Supporting Videos 4-7**). A simplified model based on the spin-flip and spin-reorientation processes is illustrated in the **Supporting Information Section 4**.

## 3. Conclusion

In conclusion, we have shown the high tunability of the magnetization switching in symmetric (monolayer/monolayer and bilayer/bilayer) and asymmetric (monolayer/bilayer) orthogonally-twisted CrSBr van der Waals heterostructures by magneto-transport measurements. On the one hand, a wide variety of switching fields depending on the number and stacking geometry of the layers and the direction of the applied field, can be obtained. On the other hand, a switch between volatile and non-volatile magnetic memory at zero-field is triggered by the proper application of an external magnetic field (magnitude and direction) in the monolayer/monolayer and monolayer/bilayer orthogonally-twisted heterostructures. On the contrary, volatile states are always observed for the orthogonally-twisted bilayer/bilayer heterostructure. This is a consequence of the intrinsic spin anisotropy of this van der Waals magnet which yields to a competition between the spin-flip and spin-reorientation processes occurring in the different layers, as supported by micromagnetic simulations. Overall, our results pinpoint the role of engineering the stacking of pristine flakes with different number of layers —which, in our case, is exemplified by considering the pristine monolayer and bilayer case as building blocks— as a versatile strategy for achieving highly tuneable twisted two-dimensional magnets, thus highlighting the number of magnetic layers as a new degree of freedom when combined with the twist-angle, with potential relevance towards the miniaturization of magnetic sensors and spintronic memory devices based on twisted vdW magnets.

## 4. Experimental Section/Methods

*Van der Waals heterostructure fabrication:* Bulk CrSBr are grown by solid state reactions, following the growth and characterization procedure as reported in our previous reference [5]. Monolayers and bilayers of CrSBr, together with few-layers graphene and h-BN, are mechanically exfoliated from their bulk counterpart. The number of layers and crystallographic axis of CrSBr is determined based on their optical contrast, as previously calibrated in our reference [5]. Crystallographic axis are determined based Selected flakes are assembled deterministically employing polycarbonate and a micromanipulator and placed on top of pre-lithographed electrodes (5 nm Ti/50 nm Au on 300 nm $SiO_2$/Si). All this process is performed under inert atmosphere conditions to avoid any possible degradation under ambient conditions.

A total of 11 devices are considered in this work (denoted as MM, MB and BB, which corresponds to monolayer/monolayer, monolayer/bilayer and bilayer/bilayer, respectively). Although the exact details vary from device to device, probably due to slight field misalignments or fabrication imperfections in the twist angle or strain of the layers, the overall phenomenology is robust and consistent among devices. Details are given in the **Supporting Information Section 1**. Pristine monolayer and bilayer devices shown in **Figure 1.c,d** correspond to the devices A5 and B6 of our previous work,[5] respectively.

*Magneto-transport measurements:* Electrical transport measurements are performed in a Quantum Design PPMS-9 cryostat with a four-probe geometry (MFLI from Zurich Instruments) using an external resistance of 1 MΩ or 10 MΩ, that is, a resistance much larger than that of the sample and a current bias within 1-10 μA.[40,41] Field sweeps are performed at 200 Oe·s⁻1, rotation sweeps at 3°·s⁻¹ and temperature sweeps at 1 K·min⁻¹, unless otherwise explicitly specified.

*Micromagnetic simulations:* DFT calculations were performed with the aid of the SIESTA package[42], by employing the Generalized Gradient Approximation (GGA) within the Perdew-Burke-Ernzerhof (PBE) scheme[43] for the exchange-correlation functionals. The pseudo-atomic-orbitals (PAOs) basis comprised a double-ζ polarized set. A 10x10x1 Monkhorst-Pack (MP) k-point grid and a mesh cutoff of 1000 Ry for real-space were used for an adequate convergency. In addition, we have considered dispersion correction for improving the DFT description of the van der Waals systems,[44] yielding to a better description of the calculated interlayer distance (from 8.85 Å to 8.17 Å for the pristine bilayer) if compared with the experimental one (7.93 Å).[35] Nonetheless, we note the small effect in the interlayer exchange value (less than 1%), being the interlayer exchange values still in the order of few μeVs (within the margin of error of the method employed). For computing the lattice structure parameters, we consider U=J=0 since it is common for GGA calculations to yield lattice parameters larger than the experimental values (**Table S3**), yielding U=J=0 the most realistic result, as already employed.[45] A Hubbard U=3 and J=1 eV parameters were added to the Cr 3d orbitals as a correction to the pseudo potentials in the frame of LDA+U scheme.[46] The DFT Hamiltonian was used to parametrize the effective atomistic spin Hamiltonian as discussed in Reference [47]. Simulations of the magnetization dynamics were performed with mumax3.[37,38] Further details are developed in the **Supporting Information Section 3**.

**Supporting Information**

Supporting Information is available from the author.

**Acknowledgements**

The authors acknowledge financial support from the European Union (ERC AdG Mol-2D 788222, FET OPEN SINFONIA 964396, Horizon Widera TRILMAX, 101159646), the Spanish MCIN (2D-HETEROS PID2020-117152RB-100, co-financed by FEDER, and Excellence Unit "María de Maeztu" CEX2019-000919-M, and 2DM PID2022-137078NB-100, co-financed by FEDER), the Generalitat Valenciana (PROMETEO Program, PO FEDER Program IDIFEDER/2021/078, a Ph.D fellowship to C.B.-C and a Grisolia Ph.D fellowship to A. R. (GRISOLIAP/2021/038)) and a Severo Ochoa Ph.D. fellowship to G.M-C. (AYUD/2021/51185, co-financed by FEDER). This study forms part of the Advanced Materials program and was supported by MCIN with funding from European Union NextGenerationEU (PRTR-C17.I1) and by Generalitat Valenciana. S.M.-V. acknowledges the support from the European Commission for a Marie Sklodowska–Curie individual fellowship No. 101103355 -

SPIN-2D-LIGHT. Computations were performed on the HPC systems Cobra and Raven at the Max Planck Computing and Data Facility.

We thank Á. López-Muñoz for his constant technical support and fundamental insights. A.R. and J.F. would like to thank A. Hierro-Rodríguez and A. García-Fuente for fruitful discussions.

Received: ((will be filled in by the editorial staff))

Revised: ((will be filled in by the editorial staff))

Published online: ((will be filled in by the editorial staff))